\documentclass[12pt,letterpaper]{article}
\usepackage{amsmath,amsfonts,amsthm,amssymb,stmaryrd,relsize,bm}
\usepackage{graphicx,relsize,epic}

\theoremstyle{plain}

\numberwithin{equation}{section}

\newcommand{\pscript}{{\mathcal P}}
\newcommand{\bbar}{\overline{b}}
\newcommand{\cbar}{\overline{c}}
\newcommand{\dbar}{\overline{d}}
\newcommand{\sbar}{\overline{s}}
\newcommand{\tbar}{\overline{t}}
\newcommand{\ubar}{\overline{u}}

\newcommand{\ab}[1]{\left|#1\right|}
\newcommand{\brac}[1]{\left\{#1\right\}}
\newcommand{\paren}[1]{\left(#1\right)}
\newcommand{\sqbrac}[1]{\left[#1\right]}

\begin{document}

\title{ELEMENTARY PARTICLES\\AND THE CAUSET APPROACH TO\\DISCRETE QUANTUM GRAVITY
}
\author{S. Gudder\\ Department of Mathematics\\
University of Denver\\ Denver, Colorado 80208, U.S.A.\\
sgudder@du.edu
}
\date{}
\maketitle

\begin{abstract}
In a previous paper, the author introduced a covariant causet ($c$-causet) approach to discrete quantum gravity. A $c$-causet is a finite partially ordered set that is invariant under labeling. The invariant labeling of a $c$-causet  $x$ enables us to uniquely specify $x$ by a sequence
$\brac{s_j(x)}$, $j=0,1,2,\ldots$, of positive integers called a shell sequence of $x$. A $c$-causet $x$ describes the microscopic structure of a possible universe at a particular time step. In general, $x$ represents one of many universes in a multiverse and $x$ grows by a single element at each time step. Since early stages of a universe were probably composed of elementary particles, we propose that elementary particles can be described by simple $c$-causets. Although we do not have a rigorous theory for such a description, we present our guess as to how it might appear. The shell sequence can be applied to find theoretical masses of particles and these seem to approximately agree with known masses. We point out that the causal order provides a unification of the strong and weak forces for elementary particles and also determines the geometry of $c$-causets which describes gravity. Moreover, it appears that some particles correspond to dark matter-energy.
\end{abstract}

\section{Microscopic Picture}  
This article is based on a covariant causal set approach to discrete quantum gravity \cite{gud13,gud14,gudprep}. For background and more details, we refer the reader to \cite{hen09,sor03,sur11}. We call a finite partially ordered set a \textit{causet} and interpret the order $a<b$ in a causet $x$ to mean that $b$ is in the causal future of $a$. We denote the cardinality of a causet $x$ by $\ab{x}$. If $x$ and $y$ are causets with $\ab{y}=\ab{x}+1$ then $x$ \textit{produces} $y$ (written $x\to y$) if $y$ is obtained from $x$ by adjoining a single maximal element $a$ to $x$. If $x\to y$ we call $y$ an \textit{offspring} of $x$.

A \textit{labeling} for a causet $x$ is a bijection $\ell\colon x\to\brac{1,2,\ldots ,\ab{x}}$ such that $a,b\in x$ with $a<b$ implies that $\ell (a)<\ell (b)$. If $\ell$ is labeling for $x$, we call $x=(x,\ell )$ an $\ell$-\textit{causet}. Two $\ell$-causets $x$ and $y$ are \textit{isomorphic} if there exists a bijection
$\phi\colon x\to y$ such that $a<b$ in $x$ if and only if $\phi (a)<\phi (b)$ in $y$ and $\ell\sqbrac{\phi (a)}=\ell (a)$ for every $a\in x$. Isomorphic
$\ell$-causets are considered identical as $\ell$-causets. We say that a causet is \textit{covariant} if it has a unique labeling (up to $\ell$-causet isomorphism) and call a covariant causet a $c$-\textit{causet}. We denote the set of a $c$-causets with cardinality $n$ by $\pscript _n$ and the set of all $c$-causets by $\pscript$. It is easy to show that any $x\in\pscript$ with $\ab{x}>1$ has a unique producer and that any $x\in\pscript$ has precisely two offspring \cite{gudprep}. It follows that $\ab{\pscript _n}=2^{n-1}$, $n=1,2,\ldots\,$.

Two elements $a,b\in x$ are \textit{comparable} if $a<b$ or $b<a$. We say that $a$ is a \textit{parent} of $b$ and $b$ is a \textit{child} of $a$ if $a<b$ and there is no $c\in x$ with $a<c<b$. A \textit{path from} $a$ \textit{to} $b$ in $x$ is a sequence $a_1=a$, $a_2,\ldots a_{n-1}$, $a_n=b$ where $a_i$ is a parent of $a_{i+1}$, $i=1,\ldots ,n-1$. The \textit{height} $h(a)$ of $a\in x$ is the cardinality minus one of the longest path in $x$ that ends with $a$. If there are no such paths, then $h(a)=0$ by convention. It is shown in \cite{gudprep} that a causet $x$ is covariant if and only if $a,b\in x$ are comparable whenever $a$ and $b$ have different heights. If $x\in\pscript$ we call the sets
\begin{equation*}
S_j(x)=\brac{a\in x\colon h(a)=j}, j=0,1,2,\ldots
\end{equation*}
\textit{shells} and the sequence of integers $s_j(x)=\ab{S_j(x)}$, $j=0,1,2,\ldots$, is the \textit{shell sequence}. A $c$-causet is uniquely determined by its shell sequence and we think of $\brac{s_j(x)}$ as describing the ``shape'' or geometry of $x$ \cite{gud13,gud14}.

The tree $(\pscript ,\to )$ can be thought of as a growth model and an $x\in\pscript _n$ is a possible universe among many universes at step (time)
$n$. An instantaneous universe $x\in\pscript _n$ grows one element at a time in one of two ways. To be specific, if $x\in\pscript _n$ has shell sequence $\paren{s_0(x),s_1(x),\ldots ,s_m(x)}$, then $x$ will grow to one of its two offspring $x\to x_0$, $x\to x_1$, where $x_0$ and $x_1$ have shell sequences
\begin{align*}
&\paren{s_0(x),s_1(x),\ldots ,s_m(x)+1}\\
&\paren{s_0(x),s_1(x),\ldots ,s_m(x),1}\\
\end{align*}
respectively.
 
In the microscopic picture, we view a $c$-causet $x$ as a framework or scaffolding for a possible universe. The vertices of $x$ represent small cells that can be empty or occupied by a particle. The shell sequence that determines $x$ gives the geometry of the framework. A labeling of $x$ gives a ``birth order'' for the elements of $x$ and since $x$ is covariant, its structure is independent of birth order.

\section{Elementary Particles} 
In the previous section, $c$-causets were employed to describe possible universes within a multiverse. In the growth model
$(\pscript, \to)$ a universe grew one element at each time step which created a history for a possible evolved universe. Each history began with one vertex and its early stages contained only a few vertices. It is reasonable to assume that those early stages were inhabited by various elementary particles.We therefore propose that an elementary particle is described by a small $c$-causet. In this section, instead of thinking of a causet as a possible instantaneous universe, we will think of it as an elementary particle.

A \textit{link} in a $c$-causet $x$ is a set $\brac{a,b}\in x$ where $a$ is a parent of $b$ (order is immaterial). A $c$-causet $x$ together with its links can be thought of as a graph. A \textit{graph path} from vertex $c$ to vertex $d$ in $x$ is a sequence of distinct incident links
$\brac{c,a_1}, \brac{a_1,a_2},\brac{a_2,a_3},\ldots ,\brac{a_n,d}$. A $c$-causet $x$ is 2-\textit{connected} if for any two comparable vertices $a,b$ of $x$ there exist at least two disjoint graph paths from $a$ to $b$.

If $a\in x$ with $x$ a $c$-causet, then the pairs $(a,0), (a,1/3), (a,-1/3)$ are called \textit{preons}. We think of a preon as an indivisible particle with electric charge $0$, $1/3$ or $-1/3$. A $c$-causet is \textit{occupied} if each of its vertices is replaced by a preon. We will classify certain occupied
$c$-causets as elementary particles. For such $c$-causets the vertices will represent preons and the edges (links) will represent forces between pairs of preons. A photon is represented by an occupied $c$-causet having an even number of vertices with shell sequence $(1,1,\ldots ,1)$ and the preons are alternatively charged $1/3,-1/3$. The reason a photon has this form is that a photon should move along a geodesic in its containing universe and we have seen in a previous study that geodesics have this structure \cite{gud13,gud14}. We consider a photon to be an elementary particle.

The other elementary particles will be represented by certain occupied 2-connected $c$-causets. These elementary particles and their combinations will correspond to matter-energy. Occupied $c$-causets that are not 2-connected (2-disconnected) correspond to dark matter-energy. The following table lists the number of $c$-causets with $n$ vertices that are 2-connected and 2-disconnected, $n=1,2,\ldots ,6$. This indicates that dark matter-energy dominates over matter-energy. In this article we only consider matter-energy; specifically, elementary particles and their combinations.
\vskip .25in

{\begin{tabular}{|c|c|c|}
\hline
Number of&Number of&Number of\\
vertices&2-connected $c$-causets&2-disconnected $c$-causets\\
\hline
1&0&1\\
\hline
2&0&2\\
\hline
3&0&4\\
\hline
4&2&6\\
\hline
5&5&11\\
\hline
6&9&23\\ 
\hline\noalign{\medskip}
\multicolumn{3}{c}%
{\textbf{Table 1}}\\
\end{tabular}}
\bigskip

We now classify the elementary particles according to their representative occupied $c$-causets. This is mainly guesswork but we present some evidence for this scheme. Except for photons, we postulate that because elementary particles are forms of matter-energy, they are described by 2-connected $c$-causets. The reason for this is that 2-connected $c$-causets have multiple graph paths so there are strong connections between their preons. This enables them to stay together and be more stable. Photons are the exception but their preons are charged so they have electromagnetic attraction while the preons of dark matter-energy presumably are electrically neutral. We see from Table~1 that there are no 2-connected $c$-causets with less than four vertices. The only 2-connected $c$-causets with four vertices are the $c$-causets with shell sequences $(1,2,1)$ and $(2,2)$. We postulate that the first corresponds to the electron $e^-$ and the second to the up quark $u$. These are illustrated in Figure~1.

\begin{center}
\includegraphics{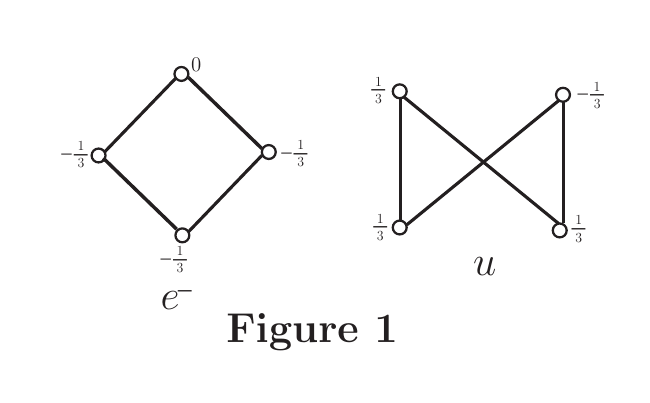}
\end{center}

In Figure~1, the vertices are preons with the labeled charges. Of course, the positron $e^+$ and the up anti-quark $\ubar$ are like $e^-$ and $u$ except the preon charges are reversed. Also, the electron neutrino $\nu _e$ is like the electron except the preon charges are zero. We propose that in the next generations, the muon and tauon leptons have the shell sequences $\mu ^-=(2,2,1)$,
$\tau ^-(3,3,4,3,1)$. The preons in $\mu ^-$ and $\tau ^-$ have charges that give the correct total charge of $-1$. As with $e^-$ and $u$, these charges are not uniquely distributed. As before, we define the antiparticles $\mu ^+$, $\tau ^+$ to be the same as $\mu ^-$, $\tau^-$ with the preon charges reversed and the neutrinos
$\nu _\mu$, $\nu _\tau$ to be the same as $\mu ^-$, $\tau ^-$ with preon charges 0.

Finally, we define the quarks $d,s,c,b,t$ as having the shell sequences, $d=(3,2)$, $s=(3,3)$, $c=(3,3,3,3)$, $b=(3,3,3,3,3,3)$, $t=(3,3,3,3,3,3,3,3)$. The preon charges are distributed to give the usual total charges $-1/3$, $-1/3$, $2/3$, $-1/3$, $2/3$, respectively. The antiquarks
$\dbar$, $\sbar$, $\cbar$, $\bbar$, $\tbar$ again have reverse preon charges. This completes our list of elementary particles. We shall not consider the Higg's boson and the weak and strong force mediating bosons here. The mesons are formed as usual from quark-antiquark pairs and the baryons are formed from quark triplets. These will be considered in the next section.

\section{Mass Formula} 

A vertex of a $c$-causet is \textit{lonesome} if it is the only vertex in its shell in $x$. For example, $e^-$ has shell sequence $(1,2,1)$ so $e^-$ has two lonesome vertices. In general, $x$ contains lonesome vertices if and only if $x$ has at least one 1 in its shell sequence. As another example, the up quark $u$ with shell sequence $(2,2)$ has no lonesome vertices. We say that a link is \textit{weak} if at least one of its two vertices is lonesome. Otherwise, a link is \textit{strong}. We associate a weak link with the weak force and a strong link with the strong force. Notice that a photon has only weak links. Since photons mediate the electromagnetic force, we see that this force is described by a sequence of an even number of weak links. In a sense, this gives a unification of these three forces of nature. Finally, the links are equivalent to the causal order which determines the geometry of $c$-causets. As we have described in \cite{gud13,gud14}, this geometry is responsible for the force of gravity. Continuing this analysis, we see that all the leptons $e^-$, $\mu ^-$, $\tau ^-$, their antiparticles and associated neutrinos all contain weak links. Except for $e^-$, $e^+$, $\nu _e$ they also contain strong links. Moreover, all the quarks have only strong links.

We now present a mass formula for the elementary particles and their compositions. Recalling that the $n$th Bohr orbit has energy proportional to $n^2$, we postulate that the part of the mass of a preon in the $n$th shell due to the weak force is also proportional to $n^2$. We next postulate that the part of the mass of a preon in the $n$th shell due to the strong force is proportional to $n^4$. Moreover, we assume that even if a particle has only strong links there are weak forces involved. To be precise, we make the following postulate where masses are in MeV.
\bigskip

\noindent\textbf{Mass Formula Postulate}: Let $p$ be a lepton, quark or hadron with shell sequence $(s_0,s_1,\ldots ,s_n)$. If $p$ has only weak links the \textit{weak-strong mass} of $p$ is
\begin{equation}         
\label{eq31}
M_p=\tfrac{1}{18}\sum _{j=0}^ns_jj^2
\end{equation}
and if $p$ has strong links the \textit{weak-strong mass} of $p$ is
\begin{equation}         
\label{eq32}
M_p=\sum _{j=0}^ns_jj^4+\tfrac{1}{18}\sum _{j=0}^ns_jj^2
\end{equation}

Of course, we do not expect $M_p$ to give the exact experimental mass because the electromagnetic force is not included. Also, this formula is unit dependent and we might get a little better agreement with experiment if we change the $1/18$ factor slightly and multiply the first summation term in \eqref{eq32} by a factor close to one.

We now apply the mass formula to the leptons and quarks. Since the electron has shell sequence $(1,2,1)$ we have
\begin{equation*}
M_e =\tfrac{1}{18}(1+8+9)=0.5\ (0.51)
\end{equation*}
where we have placed the experimental value in parenthesis. Since the muon has shell sequence $(2,2,1)$ we have
\begin{equation*}
M_\mu =2+32+81+\tfrac{1}{18}(1+8+9)\approx 116\ (105.6)
\end{equation*}
The tauon has shell sequence $(3,3,4,3,1)$ which gives
\begin{equation*}
M_\tau =3+48+324+768+625+\tfrac{1}{18}(3+12+36+48+25)\approx 1775\ (1777)
\end{equation*}
For quarks we have
\begin{align*}             
M_u&=2+32+\tfrac{1}{18}(2+8)\approx 35\\
M_d&=3+32+\tfrac{1}{18}(3+8)\approx 36\\
M_s&=3+48+\tfrac{1}{18}(3+12)\approx 52\\
M_c&=3(1+16+81+256)+\tfrac{1}{6}(1+4+9+16)\approx 1070\\
M_b&=3(1+16+81+256+625+1296)\\
  &\quad +\tfrac{1}{6}(1+4+9+16+25+36)\approx 6840\\
M_t&=3(1+16+81+256+625+1296+2401+4096)\\
  &\quad +\tfrac{1}{6}(1+4+9+1625+36+49+64)\approx 26350
\end{align*}
We do not compare these to experimental values because isolated quarks have not been found so such values are not clear..

We now consider some mesons. Until a more complete theory is established we are only guided by their quark constituency and their masses. The
$\pi ^+$-meson is composed of a $u\dbar$ pair so we assume it has the nine vertices of $u$ and $\dbar$ distributed in some order. We propose that $\pi ^+$ has the shell sequence $(5,3,1)$. This gives
\begin{equation*}
M_{\pi^+}=5+48+81+\tfrac{1}{18}(5+12+9)\approx 134\ (140)
\end{equation*}
We propose that the $\eta$-meson has the shell sequence $(2,3,3,1)$ which gives
\begin{equation*}
M_\eta =2+48+243+256+\tfrac{1}{18}(2+12+27+16)\approx 552\ (550)
\end{equation*}
We assume that the $\omega$-meson has shell sequence $(3,1,3,2)$ which gives
\begin{equation*}
M_\omega =3+16+243+512+\tfrac{1}{18}(3+4+27+32)\approx 778\ (780)
\end{equation*}
Finally, the $\eta '$-meson is composed of an $s\sbar$ pair so it has 12 vertices and we assume it has shell sequence $(6,1,2,3)$ which gives
\begin{equation*}
M_{\eta '}=6+16+162+768+\tfrac{1}{18}(6+4+18+48)\approx 956\ (958)
\end{equation*}

We next consider some baryons. The neutron $N$ is a $udd$ quark triple so we assume that $N$ has the 14 vertices of $u,d,d$ distributed in some order. We propose that $N$ has the shell sequence $(2,6,4,2)$. This gives
\begin{equation*}
M_N=2+96+324+512+\tfrac{1}{18}(2+24+36+32)\approx 939\ (940)
\end{equation*}
The proton $P$ is a $uud$ quark triple which gives 13 vertices and we assume its shell sequence is $(1,6,4,2)$. This gives
\begin{equation*}
M_P=1+96+324+512+\tfrac{1}{18}(1+24+36+32)\approx 938\ (939)
\end{equation*}
The $\Sigma ^-$-baryon is a $dds$ quark triple with 16 vertices and we assume its shell sequence is $(3,6,4,3)$. This gives
\begin{equation*}
M_{\Sigma ^-}=3+96+324+768+\tfrac{1}{18}(3+24+36+48)\approx 1197\ (1197)
\end{equation*}
The $\Xi ^-$-baryon is a $dss$ quark triple with 17 vertices and we assume its shell sequence is $(4,4,6,3)$. This gives
\begin{equation*}
M_{\Xi ^-}=4+64+486+768+\tfrac{1}{18}(4+16+54+48)\approx 1329\ (1321)
\end{equation*}
The $\Delta ^{++}$-baryon is a $uuu$ quark triple with 12 vertices and we assume its shell sequence is $(2,2,4)$. This gives
\begin{equation*}
M_{\Delta ^{++}}=4+32+162+1024+\tfrac{1}{18}(4+8+18+64)\approx 1227\ (1232)
\end{equation*}
The $\Lambda$-baryon is a $uds$ quark triple with 15 vertices and we assume its shell sequence is $(1,6,6,2)$. This gives
\begin{equation*}
M_\Lambda =1+96+486+512+\tfrac{1}{18}(1+24+54+32)\approx 1101\ (1115)
\end{equation*}
Finally, $\Omega ^-$ is an $sss$ quark triple with 18 vertices and we assume its shell sequence is $(3,4,7,4)$. This gives
\begin{equation*}
M_\Omega =3+64+567+1024+\tfrac{1}{18}(3+16+63+64)=1666\ (1672)
\end{equation*}

The author admits that this is fairly arbitrary and that one can get practically any mass he/she wants by manipulating the shell sequence. Nevertheless, it is somewhat striking that the above examples give masses so close to the experimental values. Moreover, there is a hint of a pattern in these baryon shell sequences. Notice that if there is no $s$ quark, then the last two shell numbers add to 6, if there is one $s$ quark, they add to 7 or 8, if there are two $s$ quarks, they add to 9 and if there are 3 $s$ quarks, they add to 11. It is hoped that a strengthening of this theory will point directly to the shell sequence of various particles.

We can also continue this investigation by studying atoms. For example, a hydrogen atom should be represented by a proton with the photon-electron ``kite'' of Figure~2 attached to one of its highest shell preons.

\begin{center}
\includegraphics{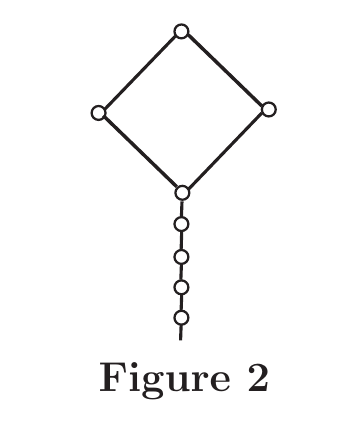}
\end{center}

\end{document}